\documentclass[preprint]{elsarticle}

\usepackage{url}
\usepackage{booktabs}
\usepackage{algpseudocode}
\usepackage{amsmath}
\usepackage{amssymb}
\usepackage{mdframed}
\usepackage{tablefootnote}
\usepackage{multicol}
\usepackage{multirow}
\usepackage{threeparttable}
\usepackage{changepage}
\usepackage{enumitem}
\usepackage[ruled, lined, linesnumbered, noend]{algorithm2e}

\setlist[description]{leftmargin=0cm, labelindent=\parindent}

\SetCommentSty{mycommfont}

\newcommand{\specialcell}[2][l]{
  \begin{tabular}[#1]{@{}l@{}}#2\end{tabular}}

\let\oldnl\nl% Store \nl in \oldnl
\newcommand{\nonl}{\renewcommand{\nl}{\let\nl\oldnl}}% Remove line number for one line

\SetKwFunction{FRecurs}{FnRecursive}%
\SetKw{Continue}{continue}

\makeatletter
\def\ps@pprintTitle{%
 \let\@oddhead\@empty
 \let\@evenhead\@empty
 \def\@oddfoot{}%
 \let\@evenfoot\@oddfoot}
\makeatother

\begin{document}
\begin{frontmatter}

\title{A General and Configurable Framework for Blockchain-based Marketplaces}

\author[uio]{Andrea Merlina\corref{mycorrespondingauthor}}
\cortext[mycorrespondingauthor]{Corresponding author}
\ead{andremer@ifi.uio.no}

\author[uio]{Roman Vitenberg}
\ead{romanvi@ifi.uio.no}

\author[uis]{Vinay Setty}
\ead{vsetty@acm.org}

\address[uio]{University of Oslo, Problemveien 7, 0315 Oslo, Norway}
\address[uis]{University of Stavanger, Kjell Arholms gate 41, 4021 Stavanger, Norway}

\begin{abstract}
The first generation of blockchain focused on digital currencies and secure storage, management and transfer of tokenized values. Thereafter, the focus has been shifting from currencies to a broader application space. 
In this paper, we systematically explore marketplace types and properties, and consider the mechanisms required to support those properties through blockchain. We propose a generic and configurable framework for blockchain-based marketplaces, and describe how popular marketplace types, price discovery policies, and other configuration parameters are implemented within the framework by presenting concrete event-based algorithms. 
Finally, we consider three use cases with widely diverging properties and show how the proposed framework supports them.
%One of the goals is to fill the gap arising when attempting to bridge digital ledgers and the trading of assets. 
%Additionally, we provide a taxonomy of marketplace properties and roles of the involved entities. 
\end{abstract}

%%Research highlights
%\begin{highlights}
%\item This work proposes a generic and configurable framework for blockchain-based marketplaces focusing on auctions while supporting different trade types.
%\item The framework provides pseudo-code for several blockchain marketplace events.
%\item We illustrate three use cases with a broad range of requirements to showcase the configurability of the framework.
%\end{highlights}

\begin{keyword}
marketplace, blockchain, framework, auction
\end{keyword}

\end{frontmatter}

\vspace{1cm}
\section{Introduction}

The online marketplaces with the most significant global net sales~\cite{EcommerceDB.10-2019} are centralized. Centralization has been a traditional design approach because it limits the overhead in the decision-making processes while providing a single reference point for national law and regulation compliance.
On the other hand, from an economic perspective, centralization goes hand in hand with monopoly, resulting in price control, unilaterally imposed commissions, and disclosure control over information~\cite{Y.Chang2019BlockchainE-Marketplace}. Besides, data integrity cannot be easily validated by external auditing, and the quality of protection against attacks is primarily up to the organization that owns the marketplace infrastructure.

%The online marketplaces with the most significant global net sales~\cite{EcommerceDB.10-2019} are centrally controlled by single authorities. Centralization has been a traditional design approach because of its simplicity, good communication complexity compared to other approaches, and limited overhead in the decision-making processes. From a regulatory perspective, it is straightforward, and laws are tuned in that sense, to manage situations where an a single organization assumes all the responsibilities. The organization is in charge of compliance with national laws and regulations.\\
%On the other hand, and from an economic perspective, centralization resonates with monopoly. Monopolies can impose price control, unilaterally decided commissions, and disclosure control over information~\cite{Y.Chang2019BlockchainE-Marketplace}. Besides, the quality of protection against attacks is primarily up to the organization that owns the marketplace infrastructure. Data integrity cannot be easily validated by external auditing. Those considerations motivate the research on different, and so far less popular, alternatives based on recent technologies.

Blockchain seems to be a natural choice as a decentralization paradigm for marketplaces, which would confer the benefits present in numerous virtual currency systems. Decisions in blockchain are inherently transparent, whereas the ledger of transactions is secure and tamper-proof. Studies~\cite{Kabi2019Blockchain-basedMarketplace,Ranganthan2018ABlockchain} indicate that, for a high volume of users, blockchain-based marketplaces are significantly cheaper than their centralized counterparts.

Unfortunately, there has been no systematic study of challenges that a blockchain-based marketplace design needs to overcome. Many marketplace policies are based on the relative order of bids in real time, which is impossible to implement over a blockchain. It is furthermore essential to define the organizational model and consider the incentives of each entity. While the exploration of blockchain-based marketplaces has already started~\cite{Banerjee2018BlockchainChallenges, Kabi2019Blockchain-basedMarketplace, Klems2017TrustlessMarketplaces, Ranganthan2018ABlockchain, Subramanian2017DecentralizedMarketplaces, Y.Chang2019BlockchainE-Marketplace}, existing proposals focus on a single specific application domain and marketplace policy, as well as a concrete blockchain system.

In this paper, we place blockchain-based marketplaces relative to other classes of marketplace implementations by considering the properties of each class. We outline the challenges and, to the best of our knowledge, propose the first generic encompassing framework for blockchain-based marketplaces. The framework solves the above challenges and supports a variety of configuration parameters at the application level (such as price discovery and bid matching policy) and the blockchain level (such as consensus protocol or blockchain type). The parameters can be set at the deployment time or dynamically on a per-trade basis.
We present concrete algorithms constituting the framework and illustrate how the framework supports three use cases with widely diverging properties: an auction of physical goods, a logo design contest and a job advertisement.
%\todo[inline]{VS: The introduction needs to be reorganized into separate paras and emphasize}
%In summary, the application of blockchains to marketplaces grants the platform some of the properties that the electronic cash system already enjoy, among which are censorship resistance and fee reduction.

%In the rest of the paper, we start by providing background information about marketplaces and the blockchain technology, illustrate their properties and introduce a general and configurable framework as part of our contributions. We proceed by presenting two uses cases based on the framework. 
%\todo[inline]{VS: If we want to save space we could omit the paper orgalinzation}

%%% about 380 words

\section{Background}
\label{ref:background}

\subsection{Background on marketplaces}
\label{ref:background_marketplace}

A marketplace is a platform that facilitates the sale of goods and services in exchange for money. Prior to the exchange, buyers and sellers advertise \textit{trading items}, place \textit{bids} for trading items, and discover a \textit{matching bid} for each item. The process of setting the item price is called price discovery~\cite{Bakos1998TheInternet}.
Common types of price discovery policies are \textit{fixed price} and \textit{auction}~\cite{McAfee1987AuctionsBidding}. In the fixed price policy, the buyer either accepts or rejects the price, similar to brick-and-mortar stores. In an auction, a well-defined negotiation phase takes place, and at its conclusion, the winning bid sets the price. 

A seller may optionally set a minimum acceptable item price for an auction, called \textit{reserve price}. A reserve price may be made public or kept secret.

%A specific feature of auctions, nevertheless optional, is the reserve price. It is the seller's imposed price that prevents any auctioned item from being sold at a less favorable price than the one set. Such a price might be established or not and could either be known to the bidders or kept secret. 
%From now on we mainly focus on auctions since fixed price markets can be trivially derived.

%\todo[inline]{VS: Is the reserve price specific to auctions only? if yes then we need to specify that, if not it needs to be rephrased. It also sounds like we only focus auctions from this point on we need to clarify that we do not focus on fixed price markets and why not}
%\todo[inline]{AM:addressed}

Within the auction policy type, commonly used price discovery policies include the following:

\begin{description}
    
    \item {\bf English auction} ---
    Buyers compete by raising their bids, which are made visible to other buyers. If the highest bid is not challenged for a certain amount of time, the item is sold to the bid proposer.
    
    \item {\bf Dutch auction} ---
    A high starting price is set by the seller and is periodically decreased using fixed decrements until either a buyer accepts it or the seller's reserve price is met. In the latter case the item remains unsold. 
   
    \item {\bf First-price sealed-bid auction} ---
    This type of auction is also known as \textit{blind auction}. Bidders propose their bid once and within a time window, without any knowledge of the other participant's offers.
    
    \item {\bf Vickrey auction}---
    Sometimes called \textit{sealed-bid second-price auction}, it is similar to the \textit{blind auction}, however the highest bidder pays the second high bid rather than its own. 
    
    \item {\bf Reverse auction}---
    In reverse auctions, the buyer publishes a request for an item or service. Sellers compete by proposing bids with decreasing prices or, differently, with increasing quality. 
    %It is not the buyer that competes proposing bids but the seller that drives down the price to provide the buyer with the item or service she needs. 
    This is commonly used in corporate procurement departments that are in need of a service while trying to optimize the contractor's cost.

\end{description}

%\todo[inline]{VS: Do we need to give any popular examples for each of these types? If they exist? Also do we need to define all these types of auctions? Are all these relevant for the paper? May be it becomes clearer after reading whole paper}

Additionally, the policies can be mixed to form hybrid versions.
%Market transactions have limited timespan and auctions in particular are strongly characterized by the duration they allow for their phases. An English auction might have a short raising-bid period in the order of few seconds, rather than a longer one. On the other hand, standard Vickrey auctions are resolved in a single wave of bids.
At the end of the price discovery phase, the matching bid is established, and the actual trading transaction may take place. A regulatory infrastructure is in place in marketplaces to deal with misbehaviors at various phases. Furthermore, a dedicated \textit{escrow} entity may be employed to resolve disputes between the trading parties.

%is followed by a flow of money between potentially multiple buyers and multiple sellers.

\subsection{Background on blockchain}
\label{ref:background_blockchain}
A blockchain is a data storage solution consisting in a sequence of blocks growing over time. Each block stores a group of immutable transactions and a hash pointer to the previous block.
A blockchain is maintained by a distributed system where a network of nodes with misaligned interests agrees on proposed transactions and securely records them. Every transaction causes a change of state in the blockchain, which is validated by the nodes, thereby providing consistency and total ordering of transactions.
%Every state change in the blockchain is approved by the validators, providing consistency and total ordering. Validators verify the proposed block and append the valid blocks to their local copy of the chain thus keeping the history of settled transactions.
%The total ordering is required to establish the validity of a bid, the precedence of transactions and in general the correctness of the marketplace.

%Smart contracts \cite{Szabo1997FormalizingHttps://ojphi.org/ojs/index.php/fm/article/view/548/469} are distributed programs that execute those state transitions. \todo[inline]{RV: do we need smart contracts?}
%\todo[inline]{AM: I think so because later on we mention that bid eval and matching is implemented in a smart contract}
The permission to become a node and the associated access policies characterize the distinction between the two different \textit{blockchain types}: permissioned and permissionless. Permissioned blockchains such as Hyperledger Fabric~\cite{hyperledger} have tight identity control, while permissionless ones such as Ethereum~\cite{Wood2014Ethereum:Ledger} allow open participation without disclosing identities. Several works have investigated permissioned and permissionless systems and their differences \cite{bessani2020byzantine, vukolic}.

Blockchain uses a \textit{consensus protocol} to maintain a consistent state among the validating nodes. 
Every state update is signed and cryptographically linked to the previous one so that any change is detectable by the nodes. This construction provides strong guarantees about integrity and authenticity.
Blockchain storage requirements grow over time, as transactions are being generated and added to the chain. To cope with the growth, some systems may perform pruning of historic transactions, deleting obsolete information.
\section{Marketplace properties}
\label{sec:marketplace-properties}

\begin{figure}
\centerline{\includegraphics[width=20pc]{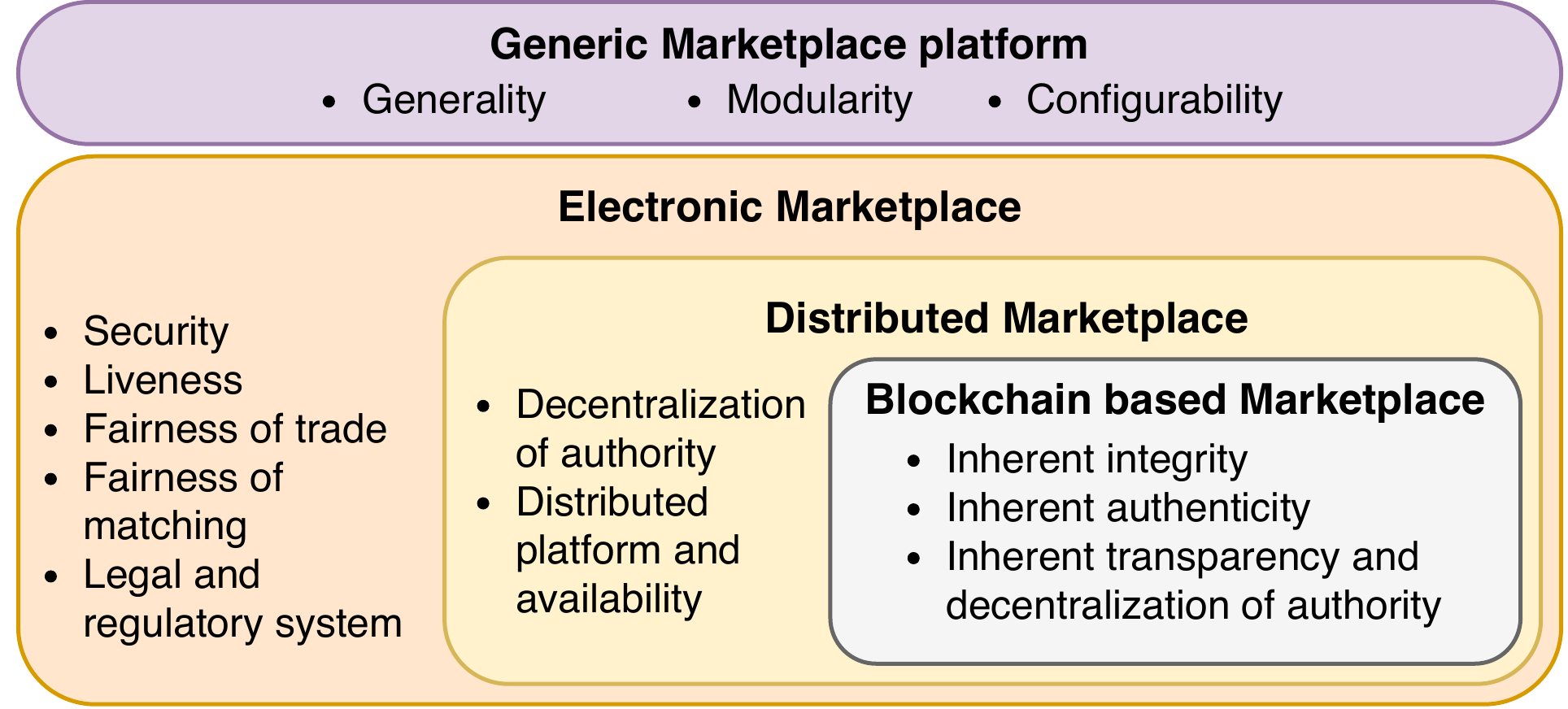}}
\vspace{-3mm}
\caption{Marketplace properties}
\label{fig:requirements}
%\vspace{-5mm}
\end{figure}

Figure~\ref{fig:requirements} illustrates properties essential for marketplaces. The properties are grouped into four categories. Electronic marketplaces may be centralized or distributed in terms of both authority and deployment. Distributed marketplaces include the properties of electronic marketplaces, while blockchain based marketplaces are a subclass of distributed ones, thereby including all of their properties.

% \begin{figure*}
% \centerline{\includegraphics[width=25pc]{images/requirements.pdf}}
% \caption{Marketplace properties}
% \label{fig:requirements}
% \end{figure*}

% ----------------------------------------------
\subsection{Properties of a generic marketplace platform}

A marketplace platform is \textit{general} if it supports various trading items, market models, and policies. %Moreover, it applies different selection algorithm for the application-level matching committee with a granularity up to the single trading item.
%The platform does not restrict its operation in vertical markets, thus saving collateral searching costs to heterogeneous buyers while offering a homogeneous level of fairness and security. Generality is supported through the properties of \textit{modularity} and \textit{configurability}. 
%Generality is challenged by technical and regulatory restrictions that introduce levels of complexity and it. is supported 
A system is \textit{modular} if its components are logically decoupled, and the implementation of each component can be easily replaced without affecting other components. For instance, a modular blockchain-based marketplace supports different consensus protocols and price discovery policies without changes to the rest of the system.
\textit{Configurability} is the capability to support a variety of configuration parameters in order to tailor the platform to specific functional and non-functional application requirements. While some parameters must be set at deployment time, others can be configured separately for each trading item or transaction, at run-time.
%Examples of parameters are trading items, metadata and duration.
%The variety of use cases in a general marketplace requires a good degree of configurability, covering multiple dimensions.

%In permissionless operation mode, buyers and sellers are not required to authenticate themselves and still the marketplace guarantees properties as security and fairness. A permissioned marketplace, on the other hand, strengthens its security assumption regarding authentication to the benefit of other properties as efficiency.
%\todo[inline]{VS: To me this definition is quite generic not sure what value it adds, while we say that modular system supports both permissioned and permissionless operations it doesnt say how or why}
%https://www.overleaf.com/3256345414cjnbppfskkzc

%\todo[inline]{VS: This is more informative than modularity.. just wondering if we need a bulletted explanation for these three items at all, can't we just summarize them in a single paragraph and emphasize these three terms.}

% ----------------------------------------------
\subsection{Electronic marketplace properties}

\begin{description}

\item {\bf Security} ---
A system is secure if it can successfully accomplish its design goals without any unintended side effects, despite impediments and threats that might subvert the system. 
An example of such an impediment would be rational behavior: an entity may deviate from the prescribed protocol for the sake of its own gain or benefit.
%Integrity and availability are implicit sub-properties in a secure system. Additionally, fraud detection and mitigation is part of the security of a marketplace.

\item {\bf Liveness} ---
A system is live if it can progress over time in its operation. For marketplace platforms, it means that associated functions, such as advertisements and bid placements, continue unimpaired. 
%A system combining security and liveness is thus able to operate securely and consistently over time.

\item {\bf Fairness of trade} ---
A marketplace is considered fair regarding the trading process if both the buyer and the seller receive the agreed trading item and the payment respectively~\cite{Banerjee2018BlockchainChallenges}. This property is violated, for example, if the buyer receives the payment yet keeps the item.
%If this does not happen, the trade has to be reverted to the starting state . 
% if the marketplace implementation is not careful this could happen

% why is it difficult to achieve? is it really a challenge in practice?
\item {\bf Fairness of matching} ---
The matching of buyer and seller is fair if winning bid selection for the trading item correctly follows the matching policies.
%no adversary subverts the winner of the trading item, for example, assigning it to a losing buyer.
This property is attained either by transparency and verifiability (when the marketplace authority is decentralized) or by trust in the organization that deploys the marketplace and performs the matching.
%In electronic marketplaces, this property holds by assuming trust and rational behavior of the marketplace, while blockchain-based marketplaces rely on computationally bounded adversaries.
%This property coupled with  \textit{Fairness of trade} guarantees correctness in the operation of the marketplace platform, the buyer and the seller.

\item {\bf Legal and regulatory system}---
Marketplaces must possess a legal infrastructure that regulates permitted behavior on the platform, thereby contributing towards correct functioning of the marketplace~\cite{Bakos1998TheInternet}. 
%A specific example is considered in the use cases.
\end{description}

% ----------------------------------------------

\subsection{Distributed marketplace properties}
\label{sec:distributed_marketplace_properties}

\begin{description}
\item {\bf Decentralization of authority}---
The control over the deployed marketplace platform is shared between a group of organizations.
For the sake of decentralization, every buyer and seller must be represented by a node in the blockchain network, either through direct ownership or a trusted proxy.
Decentralization of authority promotes transparency and decreases the risk of arbitrary decisions. 
%Additionally, misbehavior requires collusion, which makes it is less likely to happen.

\item {\bf Distributed platform and availability}---
A system is distributed if its operations are implemented by a set of components running on different machines, which orchestrate themselves by exchanging messages and appear to operate as a single process. Buyers and sellers have coherent experience of the marketplace due to this illusion of a single process.
%The appearance of being a single entity is referred to as system transparency and is necessary for buyers and sellers to have a coherent experience of the marketplace. 
%The components can be efficiently replicated, increasing fault tolerance and availability of the system.
\end{description}

% ----------------------------------------------
\subsection{Properties of blockchain-based marketplaces}
\label{sec:properties_blockchain_marketplace}

In this paper, we claim that blockchain-based marketplaces improve distributed marketplaces by inherently providing a number of desirable properties. While the same properties might potentially be attained by using alternative technologies, they are innate to blockchain.

\begin{description}
\item {\bf Inherent integrity} ---
This property is satisfied when the system is resilient to data tampering, both malicious and unwitting. Blockchain design assures such resilience through consensus and cryptographic mechanisms.

\item {\bf Inherent authenticity} ---
Authenticity refers to preventing counterfeit actions. Digital signatures and other cryptographic primitives are extensively used in blockchain design to authenticate senders of transactions and messages, thereby promoting authenticity.

\item {\bf Inherent transparency and decentralization of authority} ---
These properties lead to many benefits, including censorship resistance and fairness. Centralized marketplaces are susceptible to censorship because they are controlled by a single authority, which may unilaterally adopt policies discriminating individual buyers and sellers or preventing items or bids from being submitted to the platform.
Fairness of matching is facilitated because any observer can assess it efficiently and objectively. If the trading items are purely electronic and stored on the blockchain, fairness of trade can be easily monitored as well.
%\item{\it Stronger censorship resistance}---
%Centralized marketplaces are controlled by a single authority, thus unilaterally adopting and enforcing policies that might prevent items or bids to be submitted to the platform. A specific buyer or seller risks to be discriminated or ostracized from the marketplace. 
%
%Fairness is transparent if an observer can assess it efficiently and objectively. Inherent decentralization of authority in blockchain facilitates fairness of matching; 
\end{description}

% 740 words + 250 for the figure

%%%%%%%%%%%%%%%%%%%%%% ENTITIES AND ROLES %%%%%%%%%%%%%%%%%%%%%
\section{Marketplace model}
\label{sec:entities}

\begin{figure*}[t]
\centerline{\includegraphics[width = 30pc]{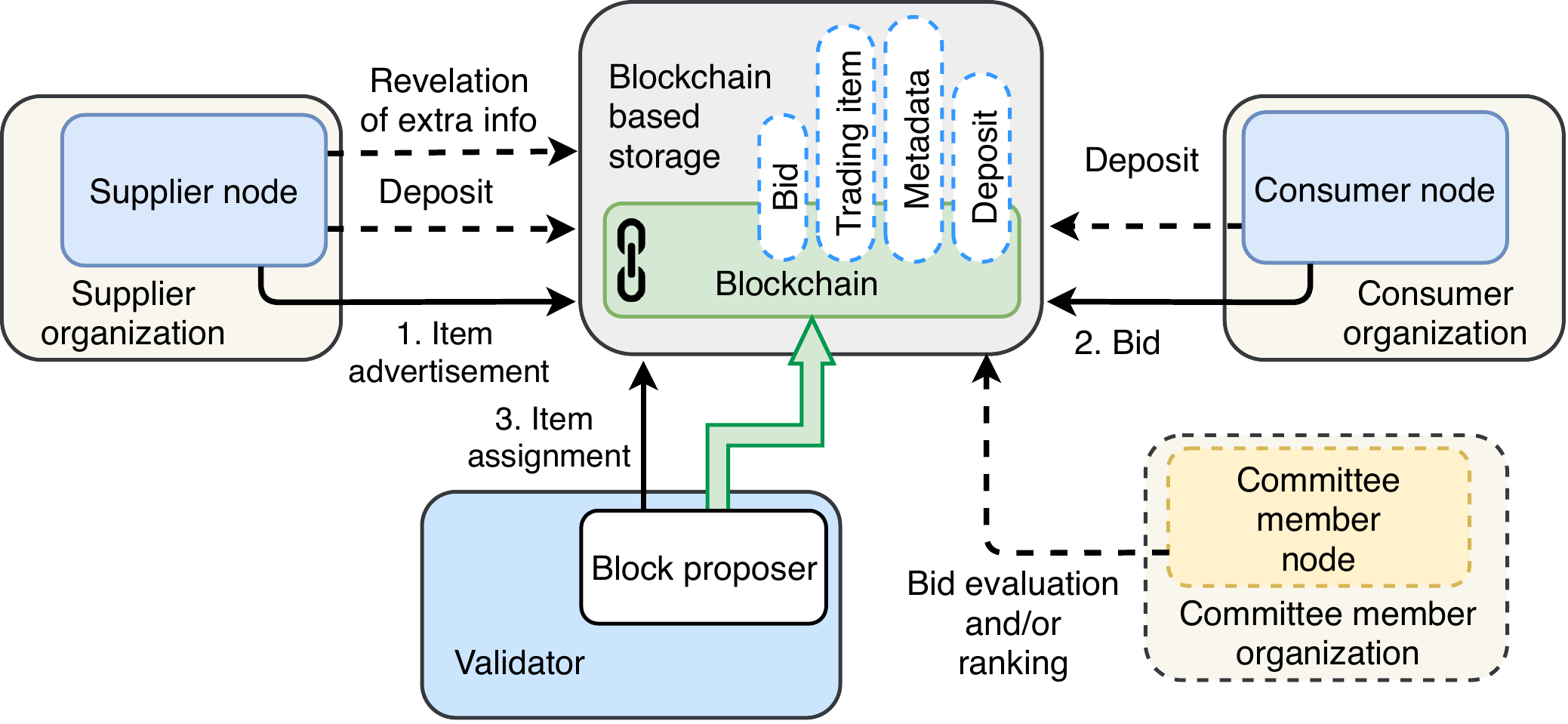}}
\vspace{-2mm}
\caption{Marketplace architecture and workflow. Dashed arrows/boxes represent optional steps/entities.}
\label{fig:workflow}
\vspace{-1mm}
\end{figure*}

%We distinguish different roles, assigned to different entities. 
At the organizational level, the system consists of nodes deploying the blockchain. The marketplace platform is implemented as a software layer atop blockchain; both permissionless and permissioned blockchains are supported, though the reliability, performance, and other nonfunctional properties might differ. Every organization or individual user must either directly run a blockchain node, or connect to a trusted blockchain node representing the user. 
%Users are represented by software clients that interact with the nodes. 
Depending on the blockchain type, there might be a dedicated consortium of organizations deploying the blockchain. Alternatively, the nodes may self-organize in a peer-to-peer fashion typical of permissionless blockchains.

Each node plays one or more roles. \textit{Suppliers} and \textit{consumers} are core marketplace roles. \textit{Escrows} and bid evaluation \textit{committees} are optional roles that may or may not be used in accordance to the requirements of the use case. At the blockchain level, every node running the marketplace and recording the blockchain state acts as a \textit{validator} for the transactions submitted to the marketplace, blocks recorded, and correctness of matching. A subset of nodes play the role of \textit{block proposers} whose task is to extend the blockchain by proposing new blocks; the subset depends on whether the blockchain is permissioned or permissionless. 

At the marketplace layer, the supplier is the role providing the trading item. The supplier is typically a seller, but it can be a buyer (of the provided service or item) in the case of reverse auctions. Consumers compete for the trading item by placing bids following the price discovery policy. 
%\rv{What does an escrow do?}
%\andrea{section 6.3 talk about escrow. maybe we can anticipate something from there?} 
A subset of all nodes may play the role of a bid evaluation committee in the context of a given trading item.

As mentioned, the bid evaluation committee is optional. In general, matching consists in evaluating individual bids and ranking them. When both evaluation and ranking follow an objective algorithm, which, e.g., selects the highest bid offered, it can directly be performed by every blockchain node so that no committee is required.
However, the matching may be based on subjective evaluation, such as that of artistic quality. When the evaluation is multi-dimensional, ranking may require subjective consideration of multiple evaluation criteria, as we exemplify in our use cases in Section~\ref{sec:use_cases}. If the evaluation and/or ranking is subjective, it is performed by a committee of experts external to the trade. While the bids are stored on blockchain, there might be additional interaction between the committee and consumer that happens off-chain. If both evaluation and ranking are subjective, the committee simply submits a blockchain transaction that declares the winning bid. If the evaluation is subjective but the ranking is objective, the committee stores evaluation of every bid on blockchain, thereby achieving a certain degree of transparency.
%Therefore, the matching is calculated objectively based on the collection of bid evaluations submitted for a trading item. 

%Besides nodes representing suppliers, consumers, escrows, and evaluators, organizations deploying the blockchain may provide additional nodes running the marketplace. 

Note that a single system may integrate the marketplace and cryptocurrency deployment by allowing transactions of both types. In fact, existence of cryptocurrency is mandatory in permissionless blockchain because of the need to pay the transaction inclusion fees and deposits. It is optional in permissioned authenticated blockchains.

\section{Challenges}
\label{sec:challenges}
A fundamental challenge that any distributed marketplace faces is the inability to establish the relative order of bids reliably. The order is essential to many trade types such as English or Dutch auctions. A centralized authority such as eBay \cite{ebay} records the order in which it receives bids. However, if the authority is decentralized, such a simple centralized solution is infeasible. In particular, the order in which bids are included into blocks in a blockchain-based marketplace does not strictly correspond to the order in which the bids were issued. We solve this challenge by replacing the strict first bid condition in all price discovery policies by a window during which bids of equal value can be proposed.
The windows are measured in the number of blocks rather than real time. They are furthermore objective because all blockchain nodes observe the same sequence of blocks. At the end of a window, the winning bid is determined by applying a pseudo-random function to the list of all bids included within that window. The application of a pseudo-random function is similar to the technique used in Algorand~\cite{algorand} for a different purpose of consensus committee selection.

Another fundamental challenge is incentivizing the entities of a blockchain-based marketplace to follow the protocol. In a permissioned blockchain that authenticates the entities, legal action can be taken against the suppliers and consumers that do not follow the rules of the trade after the matching has been completed, or against a block proposer that does not propose blocks or that includes incorrect data into block transactions. Yet, both suppliers and consumers need to pay fees to the escrow, to the bid evaluation committee (if it exists), and to organizations deploying the blockchain. The exact structure of those fees is subject to the research in economy and left for future research.

In the case of permissionless blockchain without authentication, however, additional mechanisms are required. Block proposers are incentivized by transaction inclusion fees, which may also need to be covered by suppliers and consumers. Additionally, the suppliers commit to the advertised items by placing a deposit on the blockchain, which is returned after the successful completion of the trade or if a matching bid has not been found. Similarly, the consumers place a deposit on the blockchain, which is returned after the successful completion of the trade or after the designation of another bid as matching.

\section{Solution framework}
\label{ref:architecture_overview}

% possible placement of tables 1 and 2 here

We propose a general, modular, and configurable framework that supports different market models in a single architecture. %Additionally, we illustrate how the proposed framework supports the marketplace properties listed in the previous section.
% roadmap
The framework presentation in this section is complemented by the discussion of configuration parameters in Section~\ref{sec:configurable_parameters} and of use cases in Section~\ref{sec:use_cases}.

%While  proposes new blockchain-based mechanisms for the former.

Marketplace functionality generally consists of (a) matching sellers with buyers and (b) trading. We present matching in Sections~\ref{sec:advertisement} and~\ref{sec:biddingMatchingTxSettlement} and trading in Section~\ref{sec:trading}.

The marketplace architecture and the workflow for matching are shown Figure~\ref{fig:workflow}. The workflow includes three mandatory steps: (1) Item advertisement, (2) Bid, and (3) Item assignment. Optional steps are represented with dashed arrows. 
The advertisement may include ``secret'' information, such as reserve price at an auction, which will only be revealed after all bids are placed and the sale ends.

\begin{table*}[ht!]
\begin{adjustwidth}{-2.6cm}{}
%\hspace{-4cm}
\tabcolsep 5pt
\normalsize
\centering
		
	\caption{Events generated by blockchain, suppliers, and consumers, and handled by the framework}
	\label{table:eventsAndParam}
	
	\begin{threeparttable}[t]
		\begin{tabular}{p{1.7cm} | p{2.6cm} p{1cm} p{2.9cm} | p{3.4cm} p{1.2cm} p{1.8cm}}
			
		\toprule
%		  \bf \multirow{2}{*}{Event} & 
            \bf Event &
		        \multicolumn{3}{l}{\bf Mandatory parameters} &
                \multicolumn{3}{|l}{\bf Optional parameters} \\
        		\cmidrule(lr){2-4} \cmidrule(lr){5-7}
        & {\bf Parameter} &  & {\bf Type} & {\bf Parameter} &  & {\bf Type} \\
		\midrule

	    %%%%%%%%%%% LINE 1
	    \specialcell[t]{user item\\ad. request} 
	    &
		\it{\specialcell[t]{
		tradingItem\\
		tradeType\\
		saleDuration\\
		needRevelation}}
		& 
		\it{\specialcell[t]{
		tItem \\
		tType\\
		$\delta_{\rm{sale}}$\\
		revFlag}}
		&
		{\specialcell[t]{
		domain-specific\\
		enum (see Sec.~\ref{sec:priceDiscoveryPolicy})\\
		n of blocks\\
		boolean}}
		
		&
		
		{\it{\specialcell[t]{
		startingPrice \\
		payment\\
		maxDurationNextBid\\
		maxDurationRevelation\\
		maxDurationEvaluation\\
		minimumBidIncrement\\
		reservePriceHash \\
		evalCommittee\\
		deposit}}}
		&
		{\it{\specialcell[t]{
		stPrice\\
		P\\
		$\delta_{\rm{bid}}$\\
		$\delta_{\rm{reveal}}$\\
		$\delta_{\rm{eval}}$\\
		$\Delta_{\rm{bid}}$\\
		resPHash\\
		evalCom\\
		D}}} 
		&
		{{\specialcell[t]{
		numerical \\
		cryptocur.\\
		n of blocks\\
		n of blocks\\
		n of blocks\\
		numerical \\
		numerical \\
		identity sets \\
		cryptocur.}}} 
		\\
		
		\midrule
		
		%%%%%%% LINE 2
	    \specialcell[t]{user bid \\ request} 
	    &
		\it{\specialcell[t]{
		itemAdvertisement\\
		bidContent}}
		& 
		\it{\specialcell[t]{
		A\\
		content}}
		&
		{\specialcell[t]{
		a tuple \\
		domain-specific}}
		&
		\it{\specialcell[t]{
		payment \\
		deposit}}
		&
		\it{\specialcell[t]{
		P\\
		D}} 
		&
		{\specialcell[t]{
		cryptocur. \\
		cryptocur.}} 
		\\
		
		\midrule

		%%%%%%% LINE 3
	    \specialcell[t]{block \\ inclusion\tnote{1}} 
	    &
		\it{\specialcell[t]{
		block}}
		& 
		\it{\specialcell[t]{
		B}}
		&
		\it{\specialcell[t]{
		}}
		&
		\it{\specialcell[t]{
		}}
		&
		\it{\specialcell[t]{
		}} 
		\it{\specialcell[t]{
		}} 
		\\
		
		\bottomrule   
	
	\end{tabular}
	
	\end{threeparttable}
	
	\begin{tablenotes}
	\item {\footnotesize $^{1}$  inclusion of a new block in the local copy of the blockchain}
	\end{tablenotes}

	\end{adjustwidth}
	\end{table*}

The framework is implemented as a collection of handlers for events generated by the user and blockchain. Table~\ref{table:eventsAndParam} shows the list of handled events. Item advertisement and bid proposal are  initiated by the user who deploys a particular node. New blocks, on the other hand, are received from the network and their inclusion in the local copy of the ledger triggers a multitude of derived events, such as registration of new item advertisements or bids, sale expiration, etc.

% \begin{table*}[]
%     \centering
%     \caption{List of primitives used in the framework}
%     \label{table:primitives}
%     \begin{tabular}{p{2cm} p{4cm} p{1.5cm} p{2cm}}
    
%     \toprule
%          \bf Entity & 
%          \bf Primitive &
%          \bf Input &
%          \bf Output\\
%     \toprule
%     %%% Line 1
    
%          Blockchain 
%          & 
%          \it{\specialcell[t]{
%          addTxToPropBlock\\
%          broadcastTx}}
%          & 
%          \it{\specialcell[t]{
%          Tx\\
%          Tx}}
%          &  
%          \\

%     \midrule
%     %%% Line 2
    
%          \specialcell[t]{
%          Use case\\developer }  
%          &
%          \it{\specialcell[t]{
%          customObjEval \\
%          customObjRanking\\
%          custom validation callback}} 
%          &
%          \it{\specialcell[t]{
%          bid\\ 
%          bidScores}}  
%          &
%          \it{\specialcell[t]{
%          bidScores\\
%          winningScore}}\\
         
%     \toprule     
         
%     \end{tabular}
    
%     \label{tab:my_label}
% \end{table*}

\begin{table}
    \centering
    \caption{List of primitives used in the framework}
    \label{table:primitives}
    \begin{tabular}{p{1.5cm}p{2.8cm}p{1.2cm}p{1.6cm}}
    
    \toprule
         \bf Entity & 
         \bf Primitive &
         \bf Input &
         \bf Output\\
    \toprule
    %%% Line 1
    
         Blockchain 
         & 
         \it{\specialcell[t]{
         addTxToPropBlock\\
         broadcastTx}}
         & 
         \it{\specialcell[t]{
         Tx\\
         Tx}}
         &  
         \\

    \midrule
    %%% Line 2
    
         \specialcell[t]{
         Use case\\developer }  
         &
         \it{\specialcell[t]{
         customObjEval \\
         customObjRanking\\
         %\vspace{0.5cm}
         %\rule{3pt}{2ex} 
         \addlinespace[1ex]
         custom valida-\\
         tion callbacks}} 
         &
         \it{\specialcell[t]{
         bid\\ 
         bidScores}}  
         &
         \it{\specialcell[t]{
         bidScores\\
         winningScore}}\\
         
    \toprule     
         
    \end{tabular}
    
    \label{tab:my_label}
\end{table}

Table~\ref{table:primitives} shows the primitives provided by the blockchain and use case developers that our framework utilizes. \textit{broadcastTx} disseminates a transaction to other blockchain nodes (e.g., miners in Bitcon or orderers in Hyperledger) for inclusion into a new block. In some cases, a block proposer creates a new transaction (such as an assignment of a bid to a trading item) upon processing an event. Every block proposer performs the same processing and creates the same transaction in such cases. Therefore, there is no need to disseminate the transaction throughout the network. Instead, \textit{addTxToPropBlock} adds the transaction directly to the local pool of outstanding transactions maintained by the block proposer, to be included in the next proposed block.

When the use case developer wants to employ objective evaluation of bids (see Section~\ref{sec:entities}), she needs to implement a \textit{customObjEval} function that takes a bid and returns a possibly multidimensional score for that bid. Similarly, when the use case employs objective ranking of bids, the developer needs to implement a \textit{customObjRanking} function that takes a collection of scores produced in the evaluation (by the committee or by \textit{customObjEval}) and returns the winning score.
The validation of transactions and blocks is performed collectively by the blockchain, marketplace platform, and use case developer. The blockchain validates blocks and transactions as usual. The platform validates that, e.g., deposits are present in the bid or item advertisement and that the price in the bid follows the rules. However, the use case developer may want to perform further application-specific validation, e.g., about the content of the bid. To this end, we allow the developer to define a custom validation function \textit{customValCallback} that the platform calls back.
%It is up to the use case developers to implement the appropriate validation function for the use case.  

%\vinay{Link the text to the pseudocode, for example what does validateBlock do? }
Algorithm~\ref{algo:user} shows the handling of user-initiated events, which simply consists in creating a corresponding transaction and broadcasting it to the network. The only involved element is handling of reserve price: the supplier needs to commit to a reserve price upon the advertisement and disclose it after all the bids are placed. This is achieved by supplier (a) computing a secure hash of the reserve price and including the hash in the advertisement and (b) broadcasting an additional ``revelation'' transaction after the bidding ends. Since the hash inclusion acts as a cryptographic commit, the supplier cannot modify the reserve price later without being detected. 
Algorithm~\ref{algo:main_algo} presents handling of the block inclusion event. Every node validates the block first while the rest of the handling depends on the role of the node. We present \textit{processBlockBySupplier}, \textit{processBlockByConsumer}, and \textit{processBlockByProposer} in Section~\ref{sec:marketplace_algo}. \textit{validateBlock} performs validation at the marketplace level and additionally invokes a handler that the use case developer may implement. 
We omit the code for \textit{validateBlock} due to its straightforwardness.
%Since the code for \textit{validateBlock} is relatively straightforward, it is omitted due to the space constraints.

%%%%%%%% CALLBACKS ON USER EVENTS
\begin{algorithm}[h]
\caption{Handling user-initiated events}
\label{algo:user}
\SetKwInput{Input}{Input}
\KwIn{\textit{tItem, tType, $\delta_{\rm{sale}}$, revFlag, stPrice, P, $\delta_{\rm{bid}}$, $\delta_{\rm{reveal}}$, $\Delta_{\rm{bid}}$, resPrice, evalCom, D}}

\SetKwProg{Fn}{upon}{ : }{}

%used in the use case 1
\Fn{\rm user item advertisement request}{

    \textit{resPHash} $\leftarrow$ undefined \\
    
    \If{\rm \textit{resPrice} is defined}{
        \textit{resPHash} $\leftarrow$ {compute a secure hash of \textit{resPrice} using a random salt}\\
        %commit \textit{resPrice} \tcc{Use a cryptographic primitive}
    }
    
    %\If{\rm $P$ is defined}{
    %    \textit{pTx} $\leftarrow$ \textit{signPaymentTx(P)}\\
    %}
    
    %\If{\rm $D$ is defined}{
    %    \textit{dTx} $\leftarrow$ {signDepositTx(D)}\\  
    %}
    
    %\If{\rm \textit{opt.comWList} is defined}{
    %    {evalCom} $\leftarrow$ {create list of committee %members(comWList)} \\
    %}

    %\textit{feeTx} $\leftarrow$ {s.signFeeTx(F)}\\
    
% tItem, tType, $\delta_{\rm{sale}}$, revFlag, F, stPrice, P, $\delta_{\rm{bid}}$, $\delta_{\rm{reveal}}$, $\Delta_{\rm{bid}}$, resPrice, evalCom, D    
    
    \textit{A} $\leftarrow$ create an item advertisement tx with \textit{(tItem, tType, $\delta_{\rm{sale}}$, revFlag, stPrice, P, $\delta_{\rm{bid}}$, $\delta_{\rm{reveal}}$, $\Delta_{\rm{bid}}$, resPHash, evalCom, D)} \\
    
    \textit{blockchain.broadcastTx(A)} \\
   
}

\BlankLine
\BlankLine
\KwIn{\textit{A, content, P, D}}
\Fn{\rm user bid request}{

    \tcc{In practice, we need to invoke a callback implemented by the use case developer, which would preprocess the content}
    % e.g., cryptographically commit resources
    
    %\If{\rm $C$ is defined as external digital %resource}{
    %    \textit{}{pointerC, commitC} $\leftarrow$ %{create commit and pointer to $C$}\\
    %}
    
    $bid$ $\leftarrow$ create a bid tx with (\textit{A, content, P, D})
    
    \textit{blockchain.broadcastTx(bid)} 
    
}

\end{algorithm}

%%%%%%%%%%%%%%%%%%%%%%% MAIN ALGORITHM
\begin{algorithm}[ht]

\caption{Processing new blocks by node $n$}
\label{algo:main_algo}
\SetKwProg{Fn}{upon}{ : }{}

\Fn{\rm reception of the new block to be appended in the local copy of the blockchain (block B)}{
    
    \textit{isBlockValid} $\leftarrow$ \textit{validateBlock(B)}
    
    \tcc{Block validation is split between the underlying blockchain, our platform, and use case developer, who can implement a callback}
    
    \If{$\neg$ {\rm \textit{isBlockValid}}}{
        \KwRet
    }
    \If{\rm $n$ is block proposer}{
        \textit{processBlockByProposer(B)    }
    }
    \If{\rm $n$ is supplier}{
        \textit{processBlockBySupplier(B)}
    }
    \If{\rm $n$ is consumer}{
        \textit{processBlockByConsumer(B)}
    }
}

\end{algorithm}
%%%%%%%%%%%%%%%%%%%%%%%%%%% END OF MAIN ALGORITHM

\subsection{Trade advertisement}
\label{sec:advertisement}

The trade begins with an \textit{item advertisement} in which the supplier describes the trading item and its metadata, and configures per-item parameters. The of list parameters for the item advertisement is shown in Table~\ref{table:eventsAndParam}. Mandatory parameters need to be specified for every trade type, while the combination of optional parameters varies based on the trade type. We describe the configurable parameters in depth as well as their effect on specific algorithms in Section~\ref{sec:configurable_parameters}.

Regarding the economic aspects and incentives, an item advertisement may need to include several payments in cryptocurrency: (a) a standard transaction inclusion fee for the benefit of block proposers in case of permissionless blockchain, (b) a committee fee if the evaluation committee is used, (c) an escrow fee if an escrow is used, (d) a deposit used as a collateral for faulty trade items and for not revealing reserve price in time, and (e) a payment for the services in case of reverse money flow used, e.g., in the use case in Section~\ref{sec:use_case_contest}. Each of these payments is used in a subset of scenarios and thus, optional from the platform point of view. However, they might become compulsory in a specific use case. Their presence may be enforced by a custom validation callback. As mentioned in Section
~\ref{sec:challenges}, the exact structure of the fees is left for future research in economy. However, we explicitly show how to handle the deposit and payment for the services in our algorithms.

%As part of the advertisement transaction, the supplier pays a compulsory fee which is used to incentivize the block proposers to compute and include the assignment. %\vinay{In the advertisement -> should it be when the advertisement is created or during advertisement creation?} 
%Deposits on the supplier side are optional and depend on the trade type. When present, however, they are stored on the blockchain, for the reasons described in Section \ref{sec:challenges}.

Item advertisement results in a transaction that is first validated by the validators and included in a block by the block proposer. The block is appended to the blockchain according to the consensus mechanism.

\subsection{Bidding, matching, and transaction settlement}
\label{sec:biddingMatchingTxSettlement}

Consumers bid for the advertised item according to the trade type. 
A bid transaction is accompanied by a number of optional remittances.
The list of remittances is identical to that for item advertisements,
though in each specific use case, different remittances will be compulsory
for bids. For example, the payment for services needs to be provided either
by suppliers or consumers, depending on the money flow.

%The event-based Algorithm \ref{algo:user} illustrates the passing of parameters defined by the user for both bids and advertisements. 

When the bidding phase ends, there might be an optional window of blocks that is either used by the supplier to reveal reserve price or by the committee to evaluate the bids and place the rankings on the blockchain. 
Once the bidding and the optional window end, we say that a trading item is ready for matching. At this point, we can compute the matching, assign the item to the winning bid, perform the trade, and return the deposits placed with losing bids to the consumers.

The way it is done in a blockchain-based marketplace, however, has to consider the incentives of nodes. 
The most logical way would be for each supplier and consumer to monitor the blockchain and propose relevant transactions when a trading item of interest is ready for matching: if a consumer has lost in the bidding, it will request the deposit back. A supplier and a winning consumer will request to assign the item. The disadvantage of this scheme is the proliferation of messages due to broadcasting transactions by all the entities active in the trade. We use an alternative mechanism: the block proposer itself monitors the blockchain for trading items ready for matching, computes the matching, and includes the transactions of item assignment and deposit returns into its own block. In a permissionless blockchain, the block proposer will be incentivized to perform these operations by transaction inclusion fees.

Block proposers constantly check if the specified block number for an expiration has been reached and, in the affirmative case, include an \textit{item assignment} transaction into the block that will be proposed. 
%
%Upon the item assignment, block proposers collect all the transaction and advertisement fees. 
%
The trade does not happen if there are no bids that satisfy the requirement of the matching, therefore in such cases both bids and deposits are reverted back to their owners. In the cases in which the supplier failed to include an expected revelation, the deposit of the supplier is kept as a fine. 

\subsection{Trading}
\label{sec:trading}

When the trade is entirely electronic, it is considered concluded upon the item assignment inclusion into the ledger. This applies to the use case of logo contest considered in Section~\ref{sec:use_case_contest}. However, in other cases, there is a need for an external mechanism supporting the trade of physical goods. This is commonly achieved by introducing an escrow that monitors the trade through a number of known cryptographic mechanisms extensively studied in~\cite{Goldfeder2017EscrowBitcoin}. Since the transfer of physical good is inherently centralized, it might be infeasible to replicate the delivery involvement of escrow and make it blockchain-based. However, the supplier, consumer, and escrow involved in the trade may record their steps and operations in the ledger, thereby facilitating transparency and mitigating the centralized nature of escrows.

%Since such a mechanism is not related to blockchain, we leave out of scope in this paper. 

%Disputes arising in the trade of goods are hard to settle without trust since cryptographic primitives do not provide assurance for off-chain transactions. We mitigate rational misbehavior by introducing escrows, as extensively studied in~\cite{Goldfeder2017EscrowBitcoin}. 

In case of disputes, an \textit{arbitration request} is raised either by the supplier or the consumer. The request reaches the escrow and is stored on the blockchain. The latter investigates the dispute and selects the entity to be refunded through a \textit{dispute resolution} transaction. If, on the other hand, the exchange process is finalized in its entirety without disputes, the deposit is unlocked after a safety window of blocks. In this case, by using \textit{optimistic} escrow protocols~\cite{Goldfeder2017EscrowBitcoin}, supplier and consumer can manage and spend the deposit without requiring any further action by the mediator.

%%% 1050 words + 250 for figure

%%%%%%%%%%%%%%%%%%%%%%%%%%%%%%% SUPPLIER ALGORITHM
\begin{algorithm}[ht]
\caption{Block processing by supplier $n$}
\label{algo:supplier}
% \SetKwFunction{isOddNumber}{isOddNumber}
\SetKwInput{Input}{Input}
% \SetKwInput{Output}{Output}
% \SetKwInOut{KwIn}{Input}
% \SetKwInOut{KwOut}{Output}
\KwIn{block $B$}
% \KwOut{winning bid}

\SetKwProg{Fn}{func}{ : }{}

\Fn{processBlockBySupplier}{
    
\ForEach{{\rm item assignment} $\in B$ {\rm for an item advertised by} $n$}{
    notify the user about the assignment
}
\ForEach{{\rm sale expiration for an item advertisement $A$ by $n$ due to $B$ reaching} $A.\delta_{\textrm{sale}}$}{
    \If{$A.\textit{revFlag}$}{
        \textit{Tx} $\leftarrow$ a new transaction with reserve price matching $A.\textit{resPHash}$ \\
        \tcc{In practice, the supplier would wait for an additional safety window before revealing the reserve price}
        \textit{blockchain.broadcastTx(Tx)} 
    }
}
%\ForEach{{\rm next bid expiration for an item advertisement $A$ by $n$}}{
%    \If{$A.\textit{revFlag}$}{
%        \textit{Tx} $\leftarrow$ a new transaction with reserve price matching $A.\textit{resPHash}$ \\
%        blockchain.broadcast(\textit{Tx})
%    }
%}    

}
\end{algorithm}

%%%%%%%%%%%%%%%%%%%%%%%%%%%%%%% CONSUMER  ALGORITHM
\begin{algorithm}[ht]
\caption{Block processing by consumer $n$}
\label{algo:consumer}

% \SetKwFunction{isOddNumber}{isOddNumber}
\SetKwInput{Input}{Input}
% \SetKwInput{Output}{Output}
% \SetKwInOut{KwIn}{Input}
% \SetKwInOut{KwOut}{Output}
\KwIn{block $B$}
% \KwOut{winning bid}
\SetKw{Continue}{continue}

\SetKwProg{Fn}{func}{ : }{}

\Fn{processBlockByConsumer}{

\textit{userAdvInterest} $\leftarrow$ list of item advertisements the consumer user has expressed interest for

%lowerBoundBiddingWindow $\leftarrow$ block number of either the item advertisement or the lowest block number of the highest bid(s).
%upperBoundBiddingWindow = lowerBoundBiddingWindow + $\delta_{\rm{bid}}$

\textit{blockNum} $\leftarrow$ {\# of block $B$} \\

\ForEach{A $\in$ userAdvInterest}{

    \If{\rm $B$ contains item assignment for $A$}{
        notify user about item assignment \\
        \Continue
    }
    \textit{adBlock} $\leftarrow$ block in the ledger s.t. $A\in \textit{adBlock}$ \\
    \textit{adBlockNum} $\leftarrow$ \# of \textit{adBlock} in the ledger \\

    \textit{endSaleBlockNum} $\leftarrow$ \textit{adBlockNum} $+$ $A.\delta_{\rm{sale}}$ \\
    
    %\ForEach{bid $\in$ adBids}{
        
        \If{\rm \textit{blockNum} $ > $ \textit{endSaleBlockNum}}{
            % out of the bidding window
            \Continue 
        }
        
        \textit{adBids} $\leftarrow \{\textit{bid}\, | \textit{bid}$ is for $A$ $\land$ \textit{bid} $\in B$ \} \\
        
        \eIf{adBids $\not= \emptyset$}{
        
        \If{\rm \textit{A.tType} $=$ \textit{english auction}}{
        
            notify user about $\textit{max}_{\textit{bid} \in \textit{adBids}} |\textit{bid}.P| $
            \label{algo:consumer:notifyEnglish}\\
        }
        \If{\rm \textit{A.tType} $=$ \textit{dutch auction}}{
        
            \textit{bid} $\leftarrow$ select a random bid $\in$ \textit{adBids}
            \label{algo:consumer:dutchRandom}\\
                notify user about $ | $\textit{bid.P}$ | $
        }
        
        }
        { 
         
            \If{\rm \textit{A.tType} $=$ \textit{dutch auction} }{
               $\textit{diff} \leftarrow $ (\textit{blockNum $-$ adBlockNum}) \label{algo:consumer:dutchLowerPrice:start}  \\
                \If{$\textit{diff} \mod A.\delta_{\rm{bid}} = 0$}{
                     \textit{newLowerPrice} $\leftarrow \textit{A.stPrice} - (\textit{diff} / A.\delta_{\rm{bid}}) \times A.\Delta_{\rm{bid}}$\\
                    notify user about \textit{newLowerPrice} \label{algo:consumer:dutchLowerPrice:end}
                }
            }
        }

}                
}
\end{algorithm}

%%%%%%%%%%%%%%%%%% 
\begin{algorithm}[ht!!!]
\caption{Block processing by block proposer $n$}
\label{algo:blockproposer}

\SetKwInput{Input}{Input}
\KwIn{block $B$}
\SetKwProg{Fn}{func}{ : }{}

\Fn{processBlockByProposer}{

\ForEach{\rm expiration for an item advertisement $A$ due to $B$ reaching $A.\delta_{\textrm{sale}}$ or $A.\delta_{\textrm{reveal}}$ or $A.\delta_{\textrm{eval}}$}{

\If{\rm (expiration is sale expiration) $\land$ ((\textit{A.revFlag} is true) $\lor$ (\textit{A.evalCom} is defined))}{ \label{algo:blockproposer:if}
    
        % \uIf{}{
            \Continue\\
            \tcc{Postpone the item assignment for trades that defined revelation expiration or evaluation expiration}
        % }

}

\textit{adBlock} $\leftarrow$ block in the ledger s.t. $A\in \textit{adBlock}$ \\

\textit{adBids} $\leftarrow \{\textit{bid}\, | (\textit{bid}$ is for $A$) $\land$ (\textit{bid} $\in$ a block stored in the ledger between \textit{adBlock} and $B$) \} \\

\textit{wBid} $\leftarrow$ \textit{selectWinningBid(A, adBids)} \label{ago:blockProposer:wBid}

\textit{revTx} $\leftarrow$ $\perp$

\textit{txAssignExtraFlag} $\gets$ \textit{true} \label{algo:blockProposer:flag}\\

%\textit{txAssign} \gets \textit{false}\\
%\uIf{\rm \textit{wBid} $\neq$ none}{
%            \textit{txAssign} \gets \textit{true}\\
%}

\If{\rm expiration is revelation expiration}{

        \textit{endSaleBlock} $\leftarrow$ block stored in the ledger at position \textit{adBlock} position + $A.\delta_{\rm{sale}}$ \\

        \textit{revTx} $\leftarrow$ revelation \textit{Tx} with \textit{resPrice} s.t. ((\textit{Tx} is for $A$) $\land$ (\textit{Tx} $\in$ a block stored between \textit{endSaleBlock} and $B$))
        
%        \tcc{If \textit{revTx} does not exist, we say  $\textit{revTx} = \perp$}
        
        \If{\rm (\textit{wBid} = none) $\lor$ (\textit{revTx} = $\perp$) $\lor$ (\textit{wBid.P} $<$ \textit{revTx.resPrice})}{ \label{algo:blockProposer:if2}
            
           \textit{txAssignExtraFlag} $\gets$ \textit{false} \\
        }

    }
%\If{\rm expiration is evaluation expiration}{
 
        %\tcc{Test if the committee has provided the expected evaluations. If not, apply marketplace's policy}    
    
        % \eIf{\rm \textit{wBid} $\neq$ none}{
        %     \textit{tx} $\leftarrow$ create a tx assigning \textit{A.tItem} to \textit{wBid} \\
        % }{
        %     \textit{tx} $\leftarrow$ create a ``no assignment'' tx
        % }
%    }

    \eIf{(\textit{wBid} $\neq$ none) $\land$    (\textit{txAssignExtraFlag} = \textit{true})}{
    
            \textit{tx} $\leftarrow$ create a tx assigning \textit{A.tItem} to \textit{wBid} \\
            
        }{
            \textit{tx} $\leftarrow$ create a ``no assignment'' tx 
        }
    \textit{blockchain.addTxToPropBlock(tx)}\\
    \tcc{In practice, the solution has to take into account limited block space and overflows}
    \textit{resolveMoneyFlow(A, adBids, wBid, revTx)}\\
}
 
}
\end{algorithm}

%%%%%%%%%%%%%%%%%%%%%%%%%% PARAMETERS OF ARCHITECTURE %%%%%%%%%%%%%%%%%%%%%%%%
\section{Parameters and their use in algorithms}
\label{sec:configurable_parameters}

%\rv{The following paragraph has been moved from Section 6. It needs to be properly placed in this section.}
%\textcolor{orange}{}

Our framework supports configurability at both application and system level. System level parameters were briefly described in the background on blockchain in Section~\ref{ref:background_blockchain}. 
%Note that the system level parameters 
The use case developer needs to configure the framework w.r.t. system-level parameters (the blockchain type, consensus protocol, and transaction pruning) at the deployment time rather than on a per-item basis.
%The discussion in Section~\ref{sec:use_cases} elaborates on the selection of blockchain type, consensus protocol and historical transactions pruning for each use case. 
Our framework can function atop permissioned or permissionless blockchain, however a specific use case may only be suitable for one. 
In permissioned blockchain, real entity identities are used. In the case of permissionless blockchain, identities are replaced by pseudonymous public keys. 
We now focus on marketplace-level parameters.
%Application-level parameters are described below.
%and summarized in Table~\ref{tab:parameters}.

\subsection{Trade type, bid evaluation and matching}
\label{sec:priceDiscoveryPolicy}
Table~\ref{table:eventsAndParam} shows the list of parameters for three events handled by the framework: item advertisement request, bid request and block inclusion.
The most salient parameters of an item advertisement request are the trading item \textit{tItem}, sale duration $\delta_{\rm sale}$ measured in the number of blocks, and the trade type \textit{tType}.
The framework is extensible, however it explicitly supports the trade types of \textit{English auction}, \textit{Dutch auction},  \textit{committee-based evaluation and ranking}, \textit{committee-based evaluation and custom objective ranking} and \textit{custom objective evaluation and ranking}. Note that auctions such as English and Dutch can be regarded as instances of objective evaluation and ranking due to the algorithmic selection of the winning bid, according to the deterministic and well-known matching rules. However, custom objective evaluation and ranking is a broader category, and custom policies can be implemented as pluggable algorithms by the use case developer. Algorithm~\ref{algo:winningbid} shows the selection process of the winning bid for the listed trade types.

%As an example to illustrate the different trade types, we describe in greater detail two auctions introduced in the background on marketplaces in Section~\ref{ref:background_marketplace}}.

The English auction has a fixed duration $\delta_{\rm sale}$ and a starting price \textit{stPrice}. Consumer submits bids, which must observe a minimum increment constraint $\Delta_{\rm bid}$ in order to be considered valid. Every higher bid starts a new bidding window. In absence of real time, it is possible to submit bids with the same payment \textit{P} in parallel, which may result in a single bidding window with multiple equal bids. 
%Bids offering the same price are said to be part of the same bidding window. 
The bidding phase stops due to the current block reaching the sale expiration, as described in Section~\ref{sec:biddingMatchingTxSettlement}. At that point the item becomes ready for matching. The winning bid is chosen pseudo-randomly among the bids of the last bidding window, as motivated in the challenges in Section~\ref{sec:challenges}. If the supplier establishes a reserve price by setting \textit{revFlag} to true and including the \textit{resPHash} in the advertisement, the selected bid has to be higher than the reserve price. If the supplier fails to include the revelation of the reserve price within $\delta_{\rm reveal}$ blocks from the advertisement, the item is not assigned and its deposit is kept by the block proposer as a fee. Similarly, if the selected winning bid is lower than the reserve price the item is not assigned. However, in the latter case all the deposits and payments are returned to the respective owners, with the exception of the transaction inclusion fee, which is always kept by the block proposers. 

Dutch auctions begin with a high \textit{stPrice} in the item advertisement. The price is periodically decreased, once every $\delta_{\rm bid}$ blocks, which defines a window of fixed price. The decrease is by a fixed amount defined by $\Delta_{bid}$. Within each window, consumers may bid for the item at the current price. If the window is empty of bids, the next window with a decreased price starts, unless the reserve price (if defined) is reached. In that case the auction stops without item assignment. On the other hand, if there are one or more bids in the current window, the item becomes ready for matching. In the matching, one of the bids is pseudorandomly selected as the winner, similarly to the English auction.

%Consumers take part to the auction by depositing the equivalent of the starting price. Therefore the participant list is determined at the beginning and no join is allowed during the bidding phase.

%\andrea{proposal to remove the following paragraph}
%In a contest marketplace, there is a single fixed-duration window for bids following an item advertisement. If there are submitted bids, bid evaluation starts at the end of the window. Bids are ranked either by an evaluation algorithm using objective measurable criteria, or by a bid evaluation committee within a dedicated evaluation window. The item assignment is performed based on the rankings.
%\andrea{end of proposal}

%%%%%%%%%%%%%%%%%%%%%%%%%%%%%%%%% SELECTION ON THE WINNING BID
% \newpage
\begin{algorithm}
%\setstretch{1.20} % decreases the space between lines to make the algo fit a page

\caption{Selection of the winning bid}
\label{algo:winningbid}

\SetKwInput{Input}{Input}
\SetKwInput{Output}{Output}
\KwIn{\textit{A, adBids}}
\KwOut{\textit{bid}}

\SetKwProg{Fn}{func}{ : }{}

\Fn{selectWinningBid}{
    
    \If{\textit{adBids} $= \emptyset$}{
        \KwRet none
    }
    
    \textit{adBlock} $\leftarrow$ block in the ledger s.t. $A$ $\in$ \textit{adBlock} \\

    \Switch{A.tType}{
        \Case{english auction}{
            \textit{highestPrice} $\leftarrow \textit{max}_{\textit{bid} \in \textit{adBids}} |\textit{bid}.P| $ \label{algo:wbid_english_start} \\
            \textit{candidateWinBids} $\leftarrow \{ \textit{bid} \, | \textit{bid} \in \textit{adBids} \land |\textit{bid}.P| = \textit{highestPrice} \}$ \\
            \textit{wBid} $\leftarrow$ select pseudorandom bid $\in$ \textit{candidateWinBids} \label{algo:wbid_english_end}\\
            
            %\KwRet \textit{winningBid}

        }
        \Case{auction}{

            $\textit{wBid}\!\leftarrow$ select pseudorandom bid $\in$ \textit{adBids}\\
            
            %\KwRet \textit{winningBid}

        }
        \Case{committee-based eval and ranking}{
            % job offer use case
            %\textit{adBlock} $\leftarrow$ block stored in the blockchain that includes $A$ \\

            \textit{endEvalBlock} $\leftarrow$ block stored in the ledger at position \textit{adBlock} position $+ \, A.\delta_{\rm{eval}}$ \label{algo:subjevalrank_start}
            
            \textit{evTx} $\leftarrow$ bid eval. \textit{Tx} with \textit{comDecision} s.t. ((\textit{Tx} is for $A$) $\land$ (\textit{Tx} $\in$ a block between \textit{adBlock} and \textit{endEvalBlock}))
            
            %between \textit{adBlock} and \textit{endEvalBlock}

                \If{\rm \textit{evTx} = none}{
                    \tcp{apply marketplace policy}
                    \textit{wBid} $\leftarrow$ none\\
                    
                }
                \textit{wBid} $\leftarrow$ \textit{evTx.comDecision}

            \label{algo:subjevalrank_end}

            %\KwRet \textit{winningBid}
        }
        \Case{committee-based eval and custom objective ranking}{
            % logo use case. one evaluation per bid
            %\textit{adBlock} $\leftarrow$ block stored in the blockchain that includes $A$ \\

            \textit{endEvalBlock} $\leftarrow$ block stored in the ledger at position \textit{adBlock} position $+ \, A.\delta_{\rm{eval}}$ \label{algo:objrank_start}
            
            \textit{evSet} $\leftarrow$ \{\textit{evTx} $|$ (\textit{evTx} is a bid eval. with \textit{evScore} for $A$) $\land$ (\textit{evTx} $\in$ a block between \textit{adBlock} and \textit{endEvalBlock}) \}
            
%            \tcc{Test if the committee has provided the expected evaluations. If not, apply marketplace's policy} 

                \If{\rm $|$\textit{evSet}$|$ is insufficient}{
                \tcp{apply marketplace policy}
                    \Return none
                
                }
                    \textit{bidScores} $\leftarrow \{\textit{evTx.evScore} \, | \textit{evTx} \in \textit{evSet} \}$

                    \textit{wScore} $\leftarrow$ \textit{customObjRanking(bidScores)}\\
                    
                    \textit{wBid} $\leftarrow$ select pseudorandom bid $\in$ \{\textit{b} $ |$   \textit{b} $ \in \! \textit{adBids} \land\!\textit{wScore} =\!\textit{customObjEval}(\textit{b})\}$   \\         
                    \label{algo:objrank_end}

        }
 
        \Case{custom objective eval and ranking}{
            \textit{bidScores} $\leftarrow \emptyset$ 
            
            \ForEach{bid $\in$ adBids}{
                \textit{bidScore} $\leftarrow$ \textit{customObjEval}(\textit{bid})\\
                \textit{bidScores.add(bidScore)}
            }
            
            \textit{wScore} $\leftarrow$ \textit{customObjRanking(bidScores)} \\
            
            \textit{wBid} $\leftarrow$ select pseudorandom bid $\in$ \{\textit{b} $ | $  \textit{b} $\in \! \textit{adBids} \land\!\textit{wScore} =\!\textit{customObjEval}(\textit{b})\}$
        }
        }
        
    }
    \KwRet \textit{wBid}
% }

\end{algorithm}

\subsection{Additional parameters}
Beside the evaluation and matching, the framework supports additional marketplace-level parameters. 
%In Section~\ref{sec:use_cases} we illustrate three use cases to showcase the aptness of the framework for those heterogeneous examples.

\begin{description}
\item {\bf Trading item and metadata} --- 
The item \textit{tItem} is the object of the trade. Metadata is any information that describes the item and possibly attracts interested consumers. Its format is specific to the use case.

\item {\bf Selection of the evaluation committee} ---
The evaluation committee \textit{evalCom} is a list of identities authorized to evaluate the bids. The simplest approach, and the one adopted by the framework, is to let the supplier provide such a list. In principle however, different pseudo-random selection algorithms can be used to select the evaluation committee members from a superset of designated nodes. 
The committee must submit evaluations and ranking within the duration determined by $\delta_{\rm eval}$. If the committee does not provide all required information by the expiration, the marketplace applies the relevant policy in order to determine the outcome of matching, potential reduction in the committee fee, and compensation to consumers whose bids were not sufficiently evaluated.

%The selection algorithm can have a granularity at the trading item level, or could be decided when deploying the marketplace.

\item {\bf Money and item flow} ---
In the common case, money flows from the consumer to supplier. However there are some exceptions, for instance in the case of reverse auctions and contests. 
%(see Section~\ref{ref:use_case_contest}). 
Therefore, the framework supports payments $P$ done by consumers or suppliers.
Similarly, deposits $D$ are included in bids or item advertisements. In practice, deposits are treated as locked funds that cannot be used during the locking period. 
%Validators consider invalid any transaction that transfers the deposit during the locking period.
%, preventing the owner to unduly dispose of the funds.
Upon successful item assignment, the payment is transferred to the other party either directly or via an escrow (if an escrow is used). 
Similarly, the deposit is unlocked either immediately by the platform or by the escrow at a later point.
%We do not focus on escrows, however they acts as mediators between the trading parties, which affects the described money flow.}
%\andrea{Maybe we could say that the whole trading flow that involves escrows is part of future work?}
%Additionally, in the trading phase on which we do not focus, escrow affect money flow, acting as mediator between the trading parties.} 
%For instance, the simplest scheme is one in which the consumer sends the payment to the escrow instead of the supplier. The escrow transfers the funds either to the supplier or the consumer, as it is appropriate. In this simple scheme the escrow needs to be trusted, and it mediates the money flow. More complex schemes have been proposed \cite{Goldfeder2017EscrowBitcoin}, however their study is outside the scope of this paper. 
Algorithm~\ref{algo:moneyflow} shows the resolution of the money flow in the framework.

%%%%%%%%%%%%%%%%%%%%%%%%%%%%%%%%%%% MONEY FLOW MANAGEMENT
\begin{algorithm}[ht!!!]
\caption{Money flow management}
\label{algo:moneyflow}

% \SetKwFunction{isOddNumber}{isOddNumber}
\SetKwInput{Input}{Input}
% \SetKwInput{Output}{Output}
% \SetKwInOut{KwIn}{Input}
% \SetKwInOut{KwOut}{Output}
\KwIn{\textit{A, adBids, wBid, revTx}}
% \KwOut{winning bid}

\SetKwProg{Fn}{func}{ : }{}

\Fn{resolveMoneyFlow}{

    %\textit{feeTx} $\leftarrow$ create a tx that transfers \textit{A.F} to block proposer $n$\\
    %\textit{blockchain.addTxToPropBlock(feeTx)} \\

    \If{\rm \textit{A.P} is defined}{
            \eIf{\rm \textit{wBid} $\neq$ none}{
                % there is a winning bid
                \textit{paymTx} $\leftarrow$ create a tx that transfers \textit{A.P} to the sender of \textit{wBid} (electronic trade)
                or via the escrow (physical goods) \\
            }{
                % there is no winning bid
                \textit{paymTx} $\leftarrow$ create a tx that unlocks \textit{A.P} \\
            }
            
            \textit{blockchain.addTxToPropBlock(paymTx)} \\
    }
    
    \If{\rm (\textit{A.D} is defined $\land$ \textit{A.revFlag} is false) $\lor$ (\textit{A.D} is defined $\land$ \textit{A.revFlag} is true $\land$ \textit{revTx} $\neq$ $\perp$)}{
    \tcc{The supplier loses the deposit if revTx is not included in time}
    
        \textit{depTx} $\leftarrow$ create a tx that unlocks \textit{A.D}
        
        \textit{blockchain.addTxToPropBlock(depTx)}
        \tcc{If the assignment is present and escrow is used, the escrow will unlock \textit{A.D} after the trade is completed}  \label{algo:moneyflow:comment}
        
    }

    \ForEach{bid $\in$ adBids}{
    
        \If{\rm \textit{bid.P} is defined}{
            \eIf{\rm \textit{bid} is not \textit{wBid}}{
                \textit{paymTx} $\leftarrow$ create a tx unlocking \textit{bid.P}
    
            }{
                \textit{paymTx} $\leftarrow$ create a transaction that transfers \textit{bid.P} to the supplier of $A$ or via the escrow
            }
            
            \textit{blockchain.addTxToPropBlock(paymTx)}   
        }
    
        \If{\rm \textit{bid.D} is defined}{
            \textit{depTx} $\leftarrow$ create a tx that unlocks \textit{bid.D}
            
            \textit{blockchain.addTxToPropBlock(depTx)}
            \tcc{See comment in line \ref{algo:moneyflow:comment}}  
        }
    }
}
\end{algorithm}

\item {\bf Duration and time} ---
The framework introduces a number of different duration parameters: $\delta_{\rm sale}$, $\delta_{\rm bid}$, $\delta_{\rm reveal}$ and $\delta_{\rm eval}$ for placing all the bids, placing bids for a given price, revelation of secret information by the supplier, and committee evaluation, respectively.  
As explained in Section~\ref{sec:challenges}, durations are measured in the number of blocks rather than actual time. 
%It is difficult to implement other measures such as maximum number of bids or minimum number of evaluation rankings in a blockchain-based marketplace because of diverging views in absence of centralized authority. 
The duration parameters is either set upon item advertisement or fixed upon marketplace deployment.
%Table~\ref{table:eventsAndParam} shows time-related parameters.

\end{description}

%Algorithms for handling new blocks
\subsection{Algorithms for handling new blocks}
\label{sec:marketplace_algo}

We now describe how suppliers, consumers and block proposers act upon the reception of a new block is described in the algorithms. In Algorithm~\ref{algo:supplier}, suppliers are notified about the assignments of their own advertisements. If the trade requires disclosure of information such as the reserve price, it is done upon sale expiration.  
Algorithm~\ref{algo:consumer} defines a list of advertisements the consumer is interested in and notifies the user in case of assignments. Additionally, if the bidding period has not expired yet, the user is notified about the highest bid value in English auctions (line~\ref{algo:consumer:notifyEnglish}). All bids placed during a Dutch auction will have the same value. We notify the user about the value of one such bid in line~\ref{algo:consumer:dutchRandom}. In the absence of bids, the user is notified every time the price is lowered (lines~\ref{algo:consumer:dutchLowerPrice:start} to~\ref{algo:consumer:dutchLowerPrice:end}).

The reception of a block by the block proposer is handled as shown in Algorithm~\ref{algo:blockproposer}.
Every block proposer maintains a list of unexpired trades and a counter for each trade. The counter starts from the value of the corresponding $\delta$ parameter. It is decremented upon new block inclusion. Whenever it reaches zero, an expiration event for the corresponding item advertisement is triggered.
%The event-driven processing of new blocks is shown in Algorithm \ref{algo:blockproposer}. 
Every trade undergoes either one or two phases. The first phase starts with the advertisement and ends after $\delta_{\rm sale}$ blocks. The second phase is optional and takes place when either the revelation (\textit{revFlag} is true) or the evaluation (\textit{evalCom} is defined) are employed, which are mutually exclusive within the same trade. 
%Phase two is inferred by the definition respectively of \textit{revFlag} and \textit{evalCom} contained in the item advertisement. Note that $\delta_{\rm reveal}$ and $\delta_{\rm eval}$ are strictly greater than $\delta_{\rm sale}$. 
%Therefore, in such cases, the matching is postponed at the end of phase two, as shown in line~\ref{algo:blockproposer:if}.
The winning bid is selected in line~\ref{ago:blockProposer:wBid}. With the support of an extra flag we check in line~\ref{algo:blockProposer:if2} if there is a winning bid higher than the reserve price and if the revelation transaction has been included. Based on this, the block proposer either creates a transaction assigning the item to the proposer of the winning bid, or a transaction signifying the absence of assignment. 
Algorithm results in a number of transactions being added to the pool of outstanding transactions by calling \textit{addTxToPropBlock}. Normally these transactions will be added to the next block proposed by the block proposer. A more involved algorithm is employed when some transactions are excluded from the next block because of the block size limitation; we do not present the detail because of space constraints.
%In real systems, the size of the block is limited, and therefore subject to overflow. We leave consideration on the proper balance of different transaction types inclusion as future work.  }

\section{Sketching the correctness of the solution}

The proposed framework addresses all marketplace properties presented in Section~\ref{sec:marketplace-properties}, except for the legal and regulatory system property, which must be handled by means other than a technological solution. Generality and configurability are due to the framework's ability to support a range of configuration parameters and use cases, as illustrated in Sections~\ref{sec:configurable_parameters} and~\ref{sec:use_cases}.
%First, we present the framework focusing on the support for different options regarding the configuration parameters, to demonstrate the configurability property.
%Generality, modularity, and configurability are demonstrated later on in the choice of the use cases. 
The properties of distributed and blockchain-based marketplaces described in Sections~\ref{sec:distributed_marketplace_properties} and~\ref{sec:properties_blockchain_marketplace} are acquired by construction.
We now focus on the properties of electronic marketplaces.

% what about the fact we do not justify the other properties?
The proposed solution satisfies the properties of liveness and fairness of both trade and matching.
Fairness of matching can be proven by contradiction for each of the trade type and matching policies. For example, a bid selected as winning in the English auction can never be lower than another bid. In absence of significant collusion, which is unlikely in the blockchain-based marketplace model, matching in the contest model is fair under the assumption that the decision of the evaluation committee is honest. Fairness of trade is ensured by a combination of the deposit mechanism, escrow entity, and legal obligations, though the guarantee might only be probabilistic in some cases.

The solution provides liveness if the entities correctly follow the protocol and failures/attacks do not prevent liveness of the underlying blockchain system. It can be shown for each price discovery and matching policy that the bidding protocol terminates under these assumptions and that an item assignment transaction is initiated if there is a matching bid. Besides, deposits are also returned, if previously placed.

%The system respects liveness by accomplishing the trading process without blocking.
%In the absence of disputes, the marketplace assigns the item uniquely to a winner because of the consensus protocol and the smart contract logic, which is transparent and verifiable before placing any bid. In the presence of disputes, the escrow prevents application-level deadlock by active inquiry. For the same reason, either both the buyer and the seller receive what they expect or none of them do, as dictated by the fairness of trade property.
%Fairness of matching implies that the rightful winner obtains the trading item. In a blockchain-based marketplace, the smart contract governs the matching, and tampering with this mechanism involves subverting the blockchain underlying cryptography.
%DoS and rational traders
%The proposed system is resilient to DoS attacks because of deposits that are mandatory for the transactions to be considered valid by the validators.
%\todo[inline]{VS: I thought deposits are optional? so for correctness, deposit is mandatory?}

The proposed system is resilient to, e.g., DoS attacks and rational behavior. Injection of spurious transactions is disincentivized by fees. Rational behavior of consumers and suppliers is related to the fairness of trade, as described above. Block proposers and evaluation committee members are encouraged to follow the protocol by a combination of fees, blockchain transparency, and potential legal actions.

%profit-maximizing traders. Based on the system's trust assumptions, we identify two principal approaches. The entities participating in a permissioned blockchain, are authenticated and authorized, thus legally liable for their actions. A dishonest participant can be excluded from being a member
%On the other hand, suppliers and consumers that do not trust each other are required to deposit money in escrow. DoS is limited by fees, while the economic loss that a rational trader would incur, counteracts deviations from the protocol.
	
% words 45 + 250 
\section{Supported use cases and limitations} 
\label{sec:use_cases}
% \vinay{Read Section \label{sec:use_cases} 9}

To demonstrate the generality of our framework, we present three supported use cases: an auction of physical goods, a logo design contest, and a job posting. While we have performed a similar design analysis for a real estate marketplace and the assignment process in scientific peer reviewing, we focus on the three selected use cases because of their popularity and significantly diverging requirements.
For each use case, we describe the domain, sketch the requirements and roles, and envisioned parameter configuration at the application and system level. %Table~\ref{tab:results} summarizes the three use cases and their application parameters.

%%%%%%%%%%%%%%% EBAY %%%%%%%%%%%%%%%%%
\subsection{Auction of physical goods}
The first use case is an online marketplace where multiple buyers and sellers trade items, in a fashion similar to eBay. 
%\textcolor{orange}{We discuss up to the conclusion of the matching, without detailing the following physical exchange. }
The marketplace-level parameters are as follows.

%Assessing fraudulent claims on events happening outside the scope of the blockchain is challenging. Therefore, a third party escrow intervenes in case of disputes. The escrow organization owns a node in the system and detects arbitration requests submitted to the marketplace either by the consumer or the supplier. 
%\vinay{Escrow is mentioned here, it's a bit odd to start the usecase with the dispute resolution.}
%\andrea{I agree, probably after the meeting we are removing that part.}
%
\begin{description}
\item {\bf Trading item and metadata} --- 
The \textit{tItem} is generic physical goods, for instance a hardcopy book or a domestic appliance. Metadata may include item condition, shipping information and return policies.

\item {\bf Trade type} ---
The marketplace runs instances of English auctions where suppliers decide on a per-item basis whether to set the \textit{revFlag} parameter and the value of the reserve price. Suppliers independently set a starting price \textit{stPrice} and a minimum increment $\Delta_{\rm bid}$ for the raising bids. Presence of $\Delta_{\rm bid}$ reduces the number of bids and thus, transactions.

\item {\bf Bid evaluation and matching} ---
Bid evaluation is absent since the bids can be objectively evaluated based on their proposed payment. The matching is objective and implemented according to the the English auction, as shown in Algorithm~\ref{algo:winningbid} in lines~\ref{algo:wbid_english_start} to~\ref{algo:wbid_english_end}.

%\item {\bf Selection of the evaluation committee} ---
%Not present for the reasons described before.

\item {\bf Money and item flow} ---
The item advertisement includes a fee for the block proposer, a fee for the escrow, and a deposit $D$. The bid includes a fee for the block proposer and escrow, and a payment \textit{P}. Upon the assignment inclusion in the ledger, the \textit{tItem} is ready to be shipped to the consumer that placed the winning bid. The shipment is monitored by the escrow. As part of the trade, $D$ is unlocked and $P$ is transferred to the supplier. For losing bids, on the other hand, the deposits and payments are immediately returned to the respective owners.
% say who pays the fee and provides the item

\item {\bf Duration and time} ---
Auctions on eBay typically last from one to ten days, which corresponds to roughly a few hundreds blocks at Bitcoin’s proposal rate. Therefore $\delta_{\rm sale}$ assumes a value in that range. The revelation duration $\delta_{\rm reveal}$ is set to a small number of blocks, which should provide enough time for the supplier to broadcast the revelation transaction. We do not show it in the pseudo-code, however $\delta_{\rm bid}$ may be used as an optimization for English auctions, to trigger the end of bidding before $\delta_{\rm sale}$ expires. However, the selection of the winning bid in lines~\ref{algo:wbid_english_start} to~\ref{algo:wbid_english_end} in Algorithm~\ref{algo:winningbid} remains unaltered. On the other hand $\delta_{\rm eval}$ is not defined for this use case.

\item {\bf Trading process} ---
The trade itself happens through an escrow as described in Section~\ref{sec:trading}.
\end{description}

%Note that the application parameters discussed so far are system-level independent, although specific requirements steer the design choices. 
% MIDDLEGROUND PROBLEM HERE. NOTIONS ARE JUST THROWN TO THE FACE OF THE READER WITHOUT JUSTIFICATION
Regarding system-level parameters, the marketplace is implemented as the application layer atop a permissionless blockchain such as Ethereum, allowing validators to freely join and leave.  
%No confidential information is exchanged.
Consensus protocol must support a certain transaction rate as well as a number of concurrent active users. The processing at the marketplace layer is lightweight so that the blockchain consensus protocol may become a bottleneck. 
%However, the throughput at the application layer is capped by the number of transactions the blockchain supports. 
For example, Ethereum allows support of roughly up to 20 transactions per seconds~\cite{BlockchainandScalability}. The participants, and most importantly the supplier need to consider the marketplace throughput when establishing the trade duration.
The marketplace prunes transactions belonging to correctly assigned trades that are older than an age established at marketplace deployment time.

%%%%%%%%%%%%%%%%%%%%%%%%%%%%%%%%% LOGO CONTEST
\subsection{Logo design contest}
\label{sec:use_case_contest}
% domain
The second use case illustrates how the framework supports subjective evaluation when required by the trade type.
We describe a contest for the selection of a logo i.e. a graphical representation for business purposes. The artistic quality assessment for a logo design contest is an important yet non-trivial procedure. Therefore, it requires the employment of an evaluation committee.
The supplier starts by advertising the contest on the marketplace, which prompts bid submissions. 
%Bid submissions are transparently recorded and totally-ordered in the ledger. The order does not influence the ranking but may be relevant for issues orthogonal to the marketplace, i.e. to claim novelty. 
Evaluation scores are multi-dimensional and weighted according to the objective ranking algorithm.
The marketplace-level parameters are as follows.
%Every logo submission that respects the requirements must be evaluated and counted for the final ranking.

\begin{description}
\item {\bf Trading item and metadata} ---
The \textit{tItem} is the formal announcement of the contest. Metadata includes submission guidelines describing the features the committee is going to evaluate, such as logo composition, choice of colors, novelty. Metadata include additional information about submission format, desired application domain and copyright concession.

\item {\bf Trade type} ---
The trade type is \textit{committee-based evaluation and custom objective ranking}. The artistic quality of the submission is evaluated by a human committee, however the ranking is implemented as a custom objective function by the use case developer. This increases the objectivity of the matching and improves transparency by decoupling evaluation and ranking.

\item {\bf Bid evaluation and matching} ---
Evaluation criteria and features are specified in the metadata of the item advertisement. The committee submits one transaction per bid containing the numerical scores of every evaluated feature. The scores are processed by the block proposers as shown in Algorithm~\ref{algo:winningbid} in lines~\ref{algo:objrank_start} to~\ref{algo:objrank_end}. Upon evaluation expiration, block proposers compute the matching. The logo with the highest score is selected as the winner of the contest, while possible ties are resolved with a pseudo-random function.

%\vinay{Should link the algo here}

\item {\bf Selection of the evaluation committee} ---
The committee members authorized to evaluate the bids are selected during the item advertisement. Only the authorized public keys, in case of permissionless blockchain, or real identities, for permissioned, may propose evaluations.

\item {\bf Money and item flow} ---
In this case, the logo is provided as \textit{content} inside the bid. Money follows the reverse path, from supplier to consumer. The payment value $|P|$ is decided by the supplier in the item advertisement; there is no payment in the bid. $|P|$ remains constant for the duration of the trade and is transferred to the winner in the item assignment. Deposits $D$ may be locked for both supplier and consumers, to prevent fake contests and submissions. The exact structure of the deposits is subject of economic research.

\item {\bf Duration and time} ---
A plausible duration for the logo submission $\delta_{\rm sale}$ is a couple of weeks which, at Bitcoin's proposal rate, would span an interval centered at about 2000 blocks. The committee needs several week to evaluate the submissions, which would translate to several thousand blocks.  
%A sensible choice is to provide a similar amount of time for the evaluation by setting $\delta_{\rm eval}$ betwen 1000 and 3000.
Parameters $\delta_{\rm reveal}$ and $\delta_{\rm bid}$ are not defined for the logo contest.
 
%\rv{(a) delta sale is not included in delta eval. (b) Delta cannot be measured in real time. (c) Perhaps give a recommendation for delta eval? Not sure.}
%\andrea{Should be solved. Please check}

\item {\bf Trading process} ---
The trade is purely electronic. Upon the item assignment transaction, the ownership of the winning logo is transferred to the supplier while the payment is transferred to the winning consumer.
\end{description}

%system level parameters
At the system level, the logo contest marketplace is based on a permissioned blockchain, such as Hyperledger Fabric. Validators, suppliers, consumers and all the participating entities are authenticated. Update permission, i.e., the permission to broadcast new transactions is restricted to authorized members, but for the sake of transparency, no constraint is imposed on reading the ledger. 
The expected rate of bid submissions is low, therefore the choice for the consensus protocol is driven, in this case, by the number of authenticated nodes.
A small number of nodes, in the order of a few dozens, allows the use of BFT protocols such as PBFT~\cite{castro1999practical} or BFT-SMART~\cite{bftsmart}. 
%On the other hand Hyperledger Fabric scales better in size, supporting in the order of one hundred nodes.  
Contests older than a few years and not subject to legal litigation can be pruned to free the metadata storage.

%%%%%%%%%%%%%%%%%%%%%%%% THIRD USE CASE
\subsection{Job posting}
The last use case has been selected to showcase the framework support for a less traditional marketplace application: a hiring process conducted for a joint project between collaborating companies, each of them having limited trust in the others. The open position is advertised on a blockchain-based marketplace. The candidates go through the off-chain interview process, and the final ranking is registered on the blockchain. 
%sustain multiple interviews conducted by different interviewers, which might have diverging interests and opinions about the candidate. 
%requirements
%The marketplace transparently stores the evaluation of the applicants. 
The hiring choice is securely stored for auditing and legal purposes.

\begin{description}
\item {\bf Trading item and metadata} ---
The \textit{tItem} is the job position. Metadata includes information provided by the hiring companies and general requirements for the candidates. An example of the former is contract information, annual salary and benefits.

\item {\bf Trade type} ---
The job marketplace adopts \textit{committee-based evaluation and ranking} due to the acknowledged complexity of selecting the best candidate for a job position. The high number of features that are subjectively evaluated, and their non-linear combinations make the objective ranking unsuited for the use case.

%The discovery policy is framed as a generalized reverse auction, since the supplier i.e. the hiring company requires a service in the form of workforce. The generalization follows by the increased complexity in the evaluation. In this use case, the matching is not driven by lower price offers like in standard reverse auctions, but by the quality of the bid. The concept is further clarified in the next property.

\item {\bf Bid evaluation and matching} ---
Applicants bid by submitting publicly available information about their professional career and experiences. 
The evaluation takes place off-chain. Afterwards, the committee signs a single transaction containing the evaluation and ranking of the valid applications and, optionally, details on individual evaluation motivating the ranking. 
%This process creates an auditable log of the hiring evaluation procedure. 
The matching is simple and deterministic, based on the decision of the committee, as showed in Algorithm~\ref{algo:winningbid} in lines \ref{algo:subjevalrank_start} to~\ref{algo:subjevalrank_end}.

%Quantifiable properties are stored and processed algorithmically, while qualitative characteristics are assessed and evaluated by the examiners and, need to happen off-chain. 
%$A$ maps bids from each candidates and their signed evaluations to a uni-variate dimension. The uni-variate dimension, effectively produces a ranking of the applicants that is used to select the receiver of the job offer.

\item {\bf Selection of the evaluation committee} ---
The evaluation committee \textit{evalCom} is selected by the supplier in the item advertisement.

\item {\bf Money and item flow} ---
The item assignment declares the winner of the job posting, however it does not trigger a money exchange, which instead happens off-chain.
%\rv{Is there a money exchange in this case? I thought the trade means signing the contract.}
On-chain currency is used, on the other hand, to pay the fee to the evaluation committee. 
Deposits $D$ may be required for supplier and consumers, to prevent fake job posting and applications. 
%The exact structure of the deposits is subject of economic research. 

%\vinay{What is the matching algorithm $A$? Which of the 7 algorithms?}

\item {\bf Duration and time} ---
The job advertisement is active and validators accept bids for $\delta_{\rm sale}$ number of blocks. A plausible duration for the advertisement is several weeks. After the expiration of the evaluation period $\delta_{\rm eval}$, the job position is either assigned or not, based on the evaluation of the committee. This use case does not define $\delta_{\rm bid}$ and $\delta_{\rm reveal}$.
%$R$ can be seen as a generalization of the reserve function for liquidity auctions, and prevents the supplier from committing to an outcome 
%\vinay{The notations here are not linked to the notations in Table \ref{table:eventsAndParam}. Do they need to be?}
%The duration of the employment itself is part of the metadata and orthogonal to the working of the marketplace.

\item {\bf Trading process} ---
The ``trade'' consists in signing an agreement between the supplier (which will be the employer) and the winning consumer (which will be the employee). The trade does not involve monetary transactions.
\end{description}

We expect the job posting marketplace to receive a limited number of transactions per second.
However, due to the sensitive nature of the managed information, it would need to have a fine grained control on the data dissemination.
Therefore we envision the marketplace supporting job posting advertisements to be based on a permissioned blockchain, with additional features to ensure confidentiality. The participating entities are authenticated. 
%To avoid the cost of a membership reconfiguration for every job applicant, a finite set of services acts as gatherer for the application. In this way the services act as semi-static marketplace consumers.
%\rv{I do not understand the two sentences above.}
The choice of consensus protocol is similar to that of the logo contest use case.
%is implemented with a state machine replication protocol such as BFT-SMART~\cite{bftsmart}.
%\rv{Is the choice of consensus different from logo contest?}
In the committee-based evaluation and ranking, every correct trade has exactly one bid evaluation and ranking transaction. Therefore, considering the small storage requirement for the evaluations and the assignments, pruning is not implemented.

%\vinay{What is the tuple here?}
%\textit{Data confidentiality} is not provided since bids need to be public, but \textit{durability} is assured with a proper storage replication level.
%The \textit{granularity of dissemination} for advertisements on the real estate market is restricted to a few selected nodes serving a territory.

%\input{uses_case_table.tex}

\subsection{Inherent limitations of blockchain-based marketplaces}
Consistency maintenance over geographically distributed nodes introduces sensible delays in transactions submission and confirmation, therefore posing a limit on use cases that require semi real-time execution. Blockchain-based marketplaces are thus not suitable for some application scenarios such as high-frequency trading~\cite{Kaya2016High-frequencyTrading} or real-time bidding strategies.
%, since bids might not get included into the blockchain on time.
%\todo[inline]{is collusion part of our limitations?}
%In addition, dishonest entities might collude to inflate the price artificially during the discovery.

%%%% ---- 430 words + 250 table
\section{Related Work}
\label{ref:related_work}

While there exists a significant body of research on marketplaces, they mostly focus on economic aspects and application study. An early work ~\cite{Rohm2000COPS:Markets} (published before blockchains were conceptualized), analyzes the requirements and proposes ideas for implementing a modular electronic marketplace. 
%\andrea{I don't know, I liked the previous phrase: An early work~\cite{Rohm2000COPS:Markets} proposes a model for electronic markets, analyzes the requirements, and provides initial ideas towards a modular architecture, however since it temporally anticipates the paradigm shift due to blockchain, it overlooks completely the technology.}

There are a few recent works which explore the idea of blockchain-based marketplaces. 
One of the first works describes advantages of using blockchains for electronic marketplaces for various trading items~\cite{Subramanian2017DecentralizedMarketplaces}. 
Similarly~\cite{Klems2017TrustlessMarketplaces} proposes a decentralized marketplace based on Ethereum albeit the focus is on services as the traded items. In~\cite{Ranganthan2018ABlockchain, Kabi2019Blockchain-basedMarketplace}, the authors propose a marketplace based on the Ethereum blockchain and analyze gas consumption for the smart contract execution.
Blockchain-based techniques for secure data exchange and monetization of the data is proposed in~\cite{Banerjee2018BlockchainChallenges} and in  \cite{Travizano2018Wibson:Marketplace} respectively.
A blockchain-based platform for the transfer of car ownership is studied in \cite{Notheisen2017TradingBlockchain} with a focus on software engineering aspects dictated by the provided specifications.
However, these works do not consider the challenges and limitations of blockchain-based approaches, propose concrete algorithms, or explore multiple trade types. To the best of our knowledge, our work is the first to systematically consider these challenges and address them in detailed algorithms for the matching functionality of marketplaces over blockchain.  
%perform a systematic investigation of technological challenges in implementing a generic and configurable blockchain-based framework for marketplaces.
Moreover, our proposed platform is generic: it supports a significant variety of trade types through an extensive list of configurable parameters and callbacks implemented by the use case developer.
%their usage in algorithms needed to provide support for a wide variety of trade types proposed in this paper.

In terms of commercial solutions, OpenBazaar~\cite{openbazaar}, and more recently Origami~\cite{origami}, provide a decentralized infrastructure for marketplaces. The latter is implemented on top of Ethereum, therefore adopting their token standard. The design of both solutions focuses on the trading procedure as opposed to matching. For example, the support is limited to a fixed price trade type to the best of our knowledge.

%OpenBazaar is P2P while Origami considers blockchain elements by using smart contract on Ethereum to govern escrow actions.

%In the proposed framework, we demonstrate through the use cases to support a higher degree of generality.
% \vinay{What is missing is highlighting our novel contributions compared to these related works. }

%%%%%% 170 words
\section{Conclusion and future work}
\label{ref:conclusion}
% our contributions
In this paper, we proposed a generic and configurable framework for blockchain-based marketplaces. The use of blockchain endows the marketplace the inherent properties of integrity, authenticity, and transparency. However, the decentralized nature of blockchain limits a few applications, most notably the ones involving high-frequency and real-time bidding. The deployment of well studied price discovery and matching mechanisms like auctions require careful consideration of the characteristics of blockchain systems. The extension of auctions, among other mechanisms, for blockchains is therefore non-trivial. 
The proposed framework supports various price discovery and matching policies. It includes a systematic approach to the mechanisms for item advertisement, bidding and bid evaluation, and incentives for the involved parties towards a correct behavior.

% \vinay{We also need to mention the event-based algorithms we proposed.}
%We introduce a taxonomy for blockchain-based marketplaces,   

 We also presented concrete event-based algorithms constituting our solution framework. Finally, to illustrate the configurability of the framework both at the application and blockchain level, we presented three different use cases with diverging requirements.

We identify the following directions for future work:
\begin{description}
\item{\textbf{Privacy}} --- 
Addressing the privacy of the buyers and sellers and compliance to data protection regulations such as GDPR, especially for the use cases such as ``Job posting'' is essential for decentralized marketplaces. The blockchain-based solution framework presented in this paper provides essential building blocks which may be used for implementing privacy features in the future work. 
%For example, the details of the items and users can be hidden in the public ledger using encryption techniques as suggested in \cite{Subramanian2017DecentralizedMarketplaces}. \vinay{Not sure if the last sentence is correct or needed}
\item{\bf Fee structure} --- 
Correct incentives to prevent misuse of the marketplace by the participating entities such as block proposers, committees, etc. require a detailed economic study to choose a fair fee and deposits structure. Although the framework partly addresses them, a thorough analysis is beyond the scope of this paper and left for future work.
%In order for appropriate incentivization of the participating entities such as validators, evaluation committees (for subjective evaluations) etc, and preventing misuse of the marketplace by the item advertisers, it is essential to conduct  
\item{\bf Terms-of-service} --- 
Terms-of-service support agreements between the buyers, sellers, and organizations deploying the blockchain.
\item{\bf Reputation system} --- 
In any marketplace, the consumers prefer to buy from reputed suppliers. Permissioned blockchain-based marketplaces may include such reputation systems.
\end{description}

% Directions for future work include (a) addressing the privacy of buyers and sellers, (b) analyzing the optimal fee structure from the economic perspective, (c)  and (d) providing a support for reputation systems within permissioned blockchain.

%\input{algo1}

%The latter is a natural match for permissioned blockchain, but we leave the working mechanism for future study.
%From an economic perspective, other works have demonstrated how decentralized markets can cut fees, thus providing higher revenues for merchants. 
%\todo[inline]{VS: Some of these sound like they should be in limitations?}
%%% mention Trusted Execution Environments--- 

%%%%%!!! Terms of services between : buyer, seller, organizations constituting the marketplaces.
% TEE
% Fee structure

% about 130 words

\bibliographystyle{elsarticle-num}
\bibliography{local_references}
% 330 words for the references

% 150 words for biographies

\end{document}